\documentstyle[preprint,aps,graphicx]{revtex}
\tightenlines
\begin{document}
\newcommand{\bref}[1]{eq.~(\ref{#1})}
\newcommand{\be}{\begin{equation}}
\newcommand{\en}{\end{equation}}
\newcommand{\bs}{$\backslash$}
\newcommand{\us}{$\_$}

\title{Interaction between superconductive films and magnetic nanostructures}
\author{Lars Egil Helseth}
\address{Department of Physics, University of Oslo, P.O. Box 1048 Blindern, N-0316 Oslo, Norway}

\maketitle
\begin{abstract}
The interplay between magnetic and superconductive films is studied. The
generalized London equation for this system is solved, and the magnetic
fields, the energy and the interaction forces are computed. In
particular, we focus on how to manipulate vortices using 
magnetic nanostructures.  
  
\end{abstract}

\narrowtext

\newpage

\section{Introduction}
The structure of vortices in thin films was first investigated in detail by 
Pearl\cite{Pearl}. He found that such vortices interacts mainly via their 
stray field, which extends far into the nonsuperconducting medium. Later, this
theory has been extended to thin film systems with and without 
anisotropy\cite{Clem,Carneiro,Clem1}. It has also been of cnsiderable interest to
understand Josephson vortices in thin film junctions. To this end, a 
very interesting result was obtained by Kogan $et$ $al.$ \cite{Kogan}, who 
found that the field from a Josephson junction is a superposition of fields 
from Pearl vortices along the junction with a certain line density. This 
result was derived by noting that the Pearl solution is the Green's function 
for arbitrary sources.

The possibility of enhancing the critical currents by magnetic nanostructures
have been studied by many researchers\cite{Teniers,Bulaevskii}. In particular,
Bulaevskii $et$ $al.$ showed that magnetic domain structures in a magnetic film
in close contact with a superconducting film may enhance pinning of vortices,
since this gives an opportunity to pin the magnetic flux of the vortex rather
than its core\cite{Bulaevskii}. It was suggested that the pinning of vortices
in superconductor/ferromagnetic multilayers can be many times greater than the
pinning by columnar defects. 

In general, the interaction between superconductivity and magnetism has been studied
for several decades. Systems composed of magnetic and
superconductive materials are of interest not only because they are model 
systems for the interplay of competing superconducting and magnetic order 
parameters, but also because of numerous possible applications. Recently, 
the development of magnetic thin film technology has triggered a lot of 
interest in this field. Of particular importance has been the possibility of 
examining the interaction between superconductivity and magnetism in 
high-temperature
superconductors\cite{Teniers,Bulaevskii,Sonin,Gardner,Wei,Santiago,Erdin,Erdin1,Kayali,Sonin2}. 

In a recent paper we estimated the interaction between a bulk superconductor and
a Bloch wall\cite{Helseth}. It was shown that the Bloch wall shrinks on cooling through the
superconducting transition temperature. Furthermore, we suggested that if the
magnetization of the Bloch wall is strong enough, it may capture and move the
vortices.

The aim of the current paper is to study the interaction between Pearl vortices and 
magnetic nanostructures in more detail. We solve the generalized London 
equation, and compute the magnetic fields, the energy and the interaction 
forces. In the case of a movable magnetic nanostructure 
(e.g. a domain wall), one may be able to manipulate the static as well as the
dynamic behaviour of the vortices. Thus, tunable and movable nanomagnets may be 
used as vortex manipulators.

\section{The thin film system}
Consider a thin superconductive film and thin magnetic film, both located at 
z=0 with thicknesses much smaller than the penetration depth of the
superconductor, see Fig. \ref{f1}. The two are separated by a very thin oxide layer (of thickness 
t) to avoid spin diffusion and proximity effects (which may supress the 
superconductive and magnetic order parameters)\cite{Lyuksyutov}. In general, 
the current density is a sum of the supercurrents and magnetically induced 
currents, which can be expressed through the generalized London equation as
\begin{equation}
\mbox{\boldmath $\nabla \times J$} = -\frac{1}{\lambda ^{2}} \mbox{\boldmath $H$} + 
\frac{1}{\lambda ^{2} } V (\mbox{\boldmath $\rho$})\mbox{\boldmath $\hat{e}$}_{z} + 
\mbox{\boldmath $\nabla \times \nabla \times M$}_{v}\,\,\, ,
\label{LA}
\end{equation}
where $\mbox{\boldmath $J$}$ is the current density, $\lambda$ is the penetration depth, $\mbox{\boldmath $H$}$ is the 
magnetic field and $\mbox{\boldmath $\nabla$} \times \mbox{\boldmath $M$}_{v}$ 
is the magnetically induced current. The vortex is aligned in the z direction, and its
source function $V(\mbox{\boldmath $\rho$})$ is assumed to
be rotational symmetric. In the case of a Pearl vortex we may set 
$V(\rho) =(\Phi _{0} /\mu _{0}  )\delta (\rho)$, where $\Phi _{0}$ is the flux quantum and 
$\mu _{0}$ the permeability of vacuum. The requirement of rotational symmetry should be 
relaxed when dealing with certain types of anisotropic superconductors, which are known to 
have azimuthal corrections to the vortex core\cite{Hayashi}. In this case one
should also use the anisotropic version of Eq. (\ref{LA}). 
 
Recently, the Ginzburg-Landau theory was used to calculate the vortex states in
a thin mesoscopic disk in the presence of step-like external magnetic
field\cite{Milosevic}. Naturally, one may expect more accurate results from the
Ginzburg-Landau theory, in particular when the
dimensions are small. On the other hand, the generalized London theory considered here
gives significantly reduction in computational efforts, and some important points 
could be deduced without numerical analysis. Moreover, the 
inherent shortcomings of the London model can be relaxed by introducing 
corrections to the vortex core. Thus, we believe that the generalized London model 
adopted in this paper is a reasonable first approach. 

We solve the current London equation by adopting a generalization of the very useful approach of 
Refs.\cite{Kogan,Mints}. The current can only flow in a thin layer of thickness 
$d$ ($d\ll\lambda$), and it is therefore necessary to average over this thickness and 
consider the z component only
\begin{equation}
H_{z} + \lambda _{e} \left( \frac{\partial J^{s}_{y}}{\partial x}
-\frac{\partial J^{s}_{x}}{\partial y} \right) =V(\mbox{\boldmath $\rho$}) +\lambda _{e}\left( \mbox{\boldmath $\nabla \times \nabla \times M$} \right) _{z} \,\,\, ,
\label{A}
\end{equation}
where $\lambda _{e} =\lambda ^{2} /d$ is the effective penetration depth, and $\mbox{\boldmath
$M$}=d\mbox{\boldmath $M$}_{v}$.
The sheet current flowing in the thin layer is now given by $\mbox{\boldmath $J$}_{s}=d\mbox{\boldmath $J$}$.
The Maxwell equation $\mbox{\boldmath $\nabla \times H$}=\mbox{\boldmath $J$}$
gives
\begin{equation}
J_{x} = \frac{\partial H_{z}}{\partial y}
-\frac{\partial H_{y}}{\partial z},   \,\,\,\,\,\,\,\,\,
J_{y} = \frac{\partial H_{x}}{\partial z}
-\frac{\partial H_{z}}{\partial x}\,\,\, .
\end{equation}
Since all derivatives $\partial /\partial z$ are large compared to 
the tangential $\partial /\partial \mbox{\boldmath $\rho$}$, we may set
\begin{equation}
J^{s}_{x} \approx H^{-}_{y} - H^{+}_{y},   \,\,\,\,\,\,\,\,\, 
J^{s}_{y} \approx H^{+}_{x} - H^{-}_{x} \,\,\, ,  
\end{equation}
where $H^{+}_{i}$ and $H^{-}_{i}$ (i=x, y) are the components at the upper and
lower surfaces, respectively. Since the enviroments of the upper and lower
half-spaces are identical, we have $H^{+}_{i} =-H^{-}_{i}$, which results in
\begin{equation}
J^{s}_{x} \approx  -2H^{+}_{y},  \,\,\,\,\,\,\,\,\,  J^{s}_{y} \approx 
2H^{+}_{x} \,\,\, .
\end{equation} 
Here we will only consider the upper half-space. 
Using $\mbox{\boldmath $\nabla \cdot H$} =0$, Eq. (\ref{A}) becomes
\begin{equation}
H_{z} - 2\lambda _{e} \frac{\partial H_{z}}{\partial z} = 
V (\mbox{\boldmath $\rho$}) +\lambda _{e}\left( \mbox{\boldmath $\nabla \times
\nabla \times M$} \right) _{z} 
\,\,\, .
\label{LonB}
\end{equation}
We now apply the superposition
principle to separate the contributions from the vortex and the magnetic nanostructure. 
The vortex part of Eq. (\ref{LonB}) can be written as
\begin{equation}
H_{vz} - 2\lambda _{e} \frac{\partial H_{vz}}{\partial z} = 
V(\mbox{\boldmath $\rho$}) 
\,\,\, ,
\label{D}
\end{equation}
whereas the magnetic part is
\begin{equation}
H_{mz} - 2\lambda _{e} \frac{\partial H_{mz}}{\partial z} = \lambda _{e}\left( \mbox{\boldmath $\nabla \times
\nabla \times M$} \right) _{z}
\,\,\, .
\label{E}
\end{equation}

\subsection{Vortex solution}
In order to solve Eq. (\ref{D}), it is useful to note that 
$\mbox{\boldmath $\nabla \times H$}=\mbox{\boldmath $\nabla \cdot H$}=0$ 
outside the thin films. Therefore, we can introduce a scalar potential $\phi$ which
vanishes at $z\rightarrow \infty$
\begin{equation}
\phi _{v}(\mbox{\boldmath $\rho$}, z) = \frac{1}{(2\pi)^{2}}
\int_{-\infty}^{\infty} \int_{-\infty}^{\infty}  \phi( \mbox{\boldmath
$k$} ) \exp(i \mbox{\boldmath $k \cdot \rho$} -k|z|) d^{2}k\,\,\, ,
\label{F}
\end{equation}
where $k=\sqrt{k_{x}^{2} +k_{y}^{2}}$, and $\mbox{\boldmath
$H$}=\mbox{\boldmath $\nabla$} \phi$. It is helpful to note that
$H_{z}(\mbox{\boldmath $k$})=-k\phi (\mbox{\boldmath $k$})$ for the upper
half-space. 

Applying the 2D Fourier transform to Eq. (\ref{D}) we obtain 
\begin{equation}
\phi _{v} (\mbox{\boldmath $k$}) =-\frac{V(k)}{(1+2\lambda _{e}k)k}
\,\,\, .
\end{equation}
In the case that the vortex is displaced a distance $\rho _{0}$ from the origin
we may set 
\begin{equation}
\phi _{v} (\mbox{\boldmath $k$},\rho _{0} ) =-\frac{V(k) \exp(-i\mbox{\boldmath $k \cdot
\rho _{0}$})}{(1+2\lambda _{e}k)k} \,\,\, .
\label{dispv}
\end{equation}
However, for the moment we assume that the vortex is located at the origin, and
therefore the resulting scalar potential is
\begin{equation}
\phi _{v}(\mbox{\boldmath $\rho$},z) =-\frac{1}{(2\pi)^{2}}\int_{-\infty}^{\infty} 
\int_{-\infty}^{\infty} \frac{V(k) \exp(i \mbox{\boldmath $k \cdot \rho$}
-k|z|)}{k(1+2\lambda _{e}k)} d^{2}k
\,\,\, .
\end{equation}
Due to the rotational symmetry the potential is found to be
\begin{equation}
\phi _{v}(\mbox{\boldmath $\rho$},z) =-\frac{1}{2\pi }\int_{0}^{\infty} 
V(k) \frac{J_{0} (k\rho)}{1+2\lambda _{e}k} \exp(-k|z|)dk
\,\,\, ,
\end{equation}
where we have used that
\begin{equation}
\int_{0}^{2\pi} \exp(ik\rho cos\phi )d\phi =2\pi J_{0} (k\rho) \,\,\, .
\end{equation}
To obtain the magnetic field components (in the radial and z direction), we apply the following formula
\begin{equation}
\frac{d}{d\rho} J_{0} (k\rho )=-kJ_{1} (k\rho ) \,\,\, , 
\end{equation}
and find 
\begin{equation}
H_{vz} (\mbox{\boldmath $\rho$},z) =\frac{1}{2\pi }\int_{0}^{\infty} 
kV(k)\frac{J_{0} (k\rho)}{1+2\lambda _{e}k} \exp(-k|z|)dk
\,\,\, ,
\end{equation}

\begin{equation}
H_{v\rho} (\mbox{\boldmath $\rho$},z) =\frac{1}{2\pi }\int_{0}^{\infty} 
kV(k)\frac{J_{1} (k\rho)}{1+2\lambda _{e}k} \exp(-k|z|)dk
\,\,\, .
\end{equation}

In the case of a Pearl vortex the magnetic potential and fields reduce to
\begin{equation}
\phi _{v}(\mbox{\boldmath $\rho$},z) =-\frac{\Phi _{0}}{2\pi \mu _{0}  }\int_{0}^{\infty} 
\frac{J_{0} (k\rho)}{1+2\lambda _{e}k} \exp(-k|z|)dk
\,\,\, ,
\end{equation}
\begin{equation}
H_{vz} (\mbox{\boldmath $\rho$},z) =\frac{\Phi _{0}}{2\pi \mu _{0}  }\int_{0}^{\infty} 
k\frac{J_{0} (k\rho)}{1+2\lambda _{e}k} \exp(-k|z|)dk
\,\,\, ,
\end{equation}
\begin{equation}
H_{v\rho} (\mbox{\boldmath $\rho$},z) =\frac{\Phi _{0}}{2\pi \mu _{0} }\int_{0}^{\infty} 
k\frac{J_{1} (k\rho)}{1+2\lambda _{e}k} \exp(-k|z|)dk
\,\,\, .
\end{equation}

\subsection{Magnetic solution}
In general, Eq. (\ref{E}) can be solved for arbitrary magnetization distributions. 
For simplicity, we will here consider only magnetization distributions which are directed in
the z direction and have no volume charges, 
$\mbox{\boldmath $\nabla \cdot M$}=0$. Inserted in Eq. (\ref{E}) this gives 
\begin{equation}
H_{z} - 2\lambda _{e} \frac{\partial H_{z}}{\partial z} =
 -\lambda _{e}(\mbox{\boldmath $\nabla$}^{2}
\mbox{\boldmath $M$} )_{z} 
\,\,\, .
\end{equation}
The potential from the magnetic nanostructure can be obtained in the same
way as for the vortices. We will here only work out explicitly the solution for 
two-dimensional (2D) magnetic distributions, but the 1D solution is found in a similar manner. 
Applying the 2D Fourier transform to Eq. (\ref{F}) results in 
\begin{equation}
\phi _{m} (\mbox{\boldmath $k$}) =-\lambda _{e}\frac{k M(\mbox{\boldmath $k$})}{1+2\lambda
_{e}k}
\,\,\, ,
\end{equation}
which gives the following magnetic potential:
\begin{equation}
\phi _{m}(\mbox{\boldmath $\rho$},z) =-\frac{\lambda _{e}}{(2\pi )^{2}}\int_{-\infty}^{\infty} 
\int_{-\infty}^{\infty} \frac{kM(k)}{1+2\lambda _{e}k} \exp(i \mbox{\boldmath $k
\cdot \rho$})\exp(-k|z|)d^{2}k
\,\,\, .
\label{mag}
\end{equation}
In the case of a rotational symmetric magnetization distribution, we may write
\begin{equation}
\phi_{m}(\mbox{\boldmath $\rho$},z) =-\frac{\lambda _{e}}{2\pi } 
\int_{0}^{\infty} J_{0} (k\rho) \frac{k^{2} M(k)}{1+2\lambda _{e}k} \exp(-k|z|) dk
\,\,\, .
\end{equation}
Using that $\mbox{\boldmath $H$} = \mbox{\boldmath $\nabla \phi$}$, we can
calculate the radial and vertical field components:
\begin{equation}
H_{mz} (\mbox{\boldmath $\rho$},z) = \frac{\lambda _{e}}{2\pi } 
\int_{0}^{\infty} J_{0} (k\rho) \frac{k^{3}M(k)}{1+2\lambda _{e}k} \exp(-k|z|) dk
\,\,\, ,
\end{equation}
and
\begin{equation}
H_{m\rho} (\mbox{\boldmath $\rho$},z) = \frac{\lambda _{e}}{2\pi } 
\int_{0}^{\infty} J_{1} (k\rho) \frac{k^{3}M(k)}{1+2\lambda _{e}k} \exp(-k|z|) dk
\,\,\, .
\end{equation}
So far we have neglected the thickness of the oxide layer. However, to a first 
approximation, it is easy to show that the only effect of this layer is to 
shift the z coordinate, $z\rightarrow z-t$.

\subsection{Energies and forces}
We now work out the energies and forces in the case of a circular symmetric 2D 
magnetization distribution, but will later also discuss briefly a 1D distribution 
as well. Consider a vortex located a distance $\rho _{0}$ from the origin,
and a magnetization distribution centered at the origin. The three contributions
to the energy are the magnetic self-energy, the direct 
interaction energy and the energy associated with the vortex. 

The magnetic self-energy is given by
\begin{equation}
E_{m} =  -\frac{1}{2}\mu _{0} \int_{-\infty}^{\infty} \int_{-\infty}^{\infty} 
\mbox{\boldmath $M \cdot H_{m}$} d^{2}\rho \,\,\, .
\label{energy1}
\end{equation}
This expression can be transformed by using the following
formula:
\begin{equation}
\int _{-\infty}^{\infty} \int_{-\infty}^{\infty}A(\rho)B(\rho) d ^{2}\rho 
=\frac{1}{(2\pi )^{2}} \int_{-\infty}^{\infty} \int _{-\infty}^{\infty}
A(-k)B(k) d^{2}k\,\,\, .
\label{help1}
\end{equation}
Thus, in the case of a circular symmetric magnetization distribution we have
\begin{equation}
E_{1} =  -\frac{\mu _{0} \lambda _{e}}{4\pi} \int_{0}^{\infty}
k^{3} \frac{M(k)M(k)}{1+2\lambda _{e}k} dk     \,\,\, ,
\label{energy11}
\end{equation}
where we have used that M(-k)=M(k) and also that $d^{2}k \rightarrow kdk d\phi$ 
(and integrated over the azimuthal angle). 

The direct interaction energy between the vortex and the
magnetization is given by
\begin{equation}
E_{vm} = -\mu _{0} \int_{-\infty}^{\infty} \int_{-\infty}^{\infty} 
\mbox{\boldmath $M \cdot H_{v}$} d^{2}\rho \,\,\, ,
\end{equation}
where the vortex is assumed to be displaced a distance $\rho _{0}$ from the origin. Then, in 
the cylindrical symmetric situation one finds
\begin{equation}
E_{vm} =  -\frac{\mu _{0}}{2\pi} \int_{0}^{\infty} kV(k)\frac{M(k) 
J_{0} (k\rho _{0}) }{1+2\lambda _{e}k} dk     \,\,\, .
\label{energyvm}
\end{equation}
The associated force can be found by taking
the derivative with respect to $\rho _{0}$, resulting in
\begin{equation}
F_{vm} = -\frac{\mu _{0}}{2\pi} \int_{0}^{\infty} k^{2}V(k)\frac{M(k) 
J_{1} (k\rho _{0}) }{1+2\lambda _{e}k} dk     \,\,\, .
\label{forcevm}
\end{equation}

The energy associated with vortex field is expressed as
\begin{equation}
E_{3} =  \int_{V}\left( \frac{1}{2} \mu _{0} H_{v}^{2} +\frac{1}{2}
n_{s}mv_{s}^{2} \right) dV \,\,\, ,
\label{energy3}
\end{equation}
where $n_{s}$ is the density of superconducting electrons and $v_{s}$ is 
their velocity. This expression is
most easily transformed by applying the formalism of Ref. \cite{Mints}.
In particular, the energy associated with a Pearl vortex can be evaluated to
logarithmic accuracy as
\begin{equation}
E_{3} \approx \frac{\Phi _{0} ^{2}}{4\pi \mu _{0} \lambda _{e}} ln\left( \frac{4\pi
\lambda _{e}}{\xi} \right)\,\,\, ,
\end{equation}
where $\xi$ is the coherence length. More accurate expressions are obtained by
introducing the appropriate vortex source function, but this will not be considered here.

\section{Interaction with a bubble}
Consider first a magnetic bubble that can be moved, generated or annihilated by 
external stress patterns or magnetic fields\cite{Eschenfelder}. Let the vortex 
be separated a distance $\rho_{0}$ from the magnetic bubble located at the 
origin. The system has cylindrical
symmetry, and the magnetic bubble have a magnetization vector pointing in 
the z direction given by
\begin{displaymath}
\mbox{\boldmath $M$} = \left\{ \begin{array}{ll}
M_{0} \mbox{\boldmath $\hat{e}$}_{z} & \textrm{if $\rho \leq W$}\\
0 & \textrm{if $\rho > W$}\\  
\end{array} \right. .
\end{displaymath}
The Fourier transform of this distribution is
\begin{equation}
M(k)=2\pi M_{0} W \frac{J_{1} (kW)}{k} \,\,\, ,
\end{equation}
which generates the following magnetic field components:
\begin{equation}
H^{b}_{z} (\mbox{\boldmath $\rho$},z) = \lambda _{e}WM_{0} 
\int_{0}^{\infty} J_{0} (k\rho) \frac{k^{2}J_{1} (kW)}{1+2\lambda _{e}k}
\exp(-k|z|) dk \,\,\, ,
\label{bubble1}
\end{equation}
\begin{equation}
H^{b}_{\rho} (\mbox{\boldmath $\rho$},z) = \lambda _{e}WM_{0} 
\int_{0}^{\infty} J_{1} (k\rho) \frac{k^{2}J_{1} (kW)}{1+2\lambda _{e}k}
\exp(-k|z|) dk \,\,\, .
\label{bubble2}
\end{equation}
These components were also found in
Refs. \cite{Erdin1,Kayali} using a different method. To obtain the total field 
one should add the vortex contribution. The energy is found to be
\begin{equation}
E_{b} = \frac{\Phi _{0} ^{2}}{4\pi \mu _{0} \lambda _{e}} ln\left( \frac{4\pi
\lambda _{e}}{\xi} \right) - M_{0} \Phi_{0} W \int_{0}^{\infty} \frac{J_{1} (kW) 
J_{0} (k\rho _{0}) }{1+2\lambda _{e}k} dk  - \lambda _{e}\mu _{0}M_{0}^{2}\pi W^{2} \int_{0}^{\infty} 
k\frac{J_{1}^{2} (kW)}{1+2\lambda _{e}k} dk \,\,\, ,
\label{bubble-energy}
\end{equation}
and the direct force between the bubble and the vortex is given by
\begin{equation}
F_{vm} =  -M_{0} W\Phi _{0} \int_{0}^{\infty} k\frac{J_{1} (kW) 
J_{1} (k\rho _{0}) }{1+2\lambda _{e}k} dk     \,\,\, .
\label{bubble-vortex}
\end{equation}
Figure \ref{f2} shows the force when $W=\lambda_{e}$, and we see that it falls
off rather quickly. Moreover, we note that there are peaks at the maximum
magnetization gradients, as expected. A vortex in the vicinity of the bubble 
will be either repelled or attracted to the center of the 
bubble, where it eventually comes to rest. The vortex-bubble system may still
attract more vortices, but now vortex-vortex repulsion must be taken into 
account, which is outside the scope of the current study.  

Consider now an annular structure as illustrated in Fig. \ref{f3} where the 
inner dot has magnetization $M_{1}$ and the outer annulus $M_{2}$. Then the force
becomes
\begin{equation}
F_{vm} =  -M_{1} W_{1}\Phi _{0} \int_{0}^{\infty} \left[ J_{1} (kW_{1})
+\frac{M_{2}W_{3}}{M_{1}W_{1}} J_{1}(kW_{3}) -\frac{M_{2}W_{2}}{M_{1}W_{1}}
J_{1}(kW_{2}) \right] \frac{kJ_{1} (k\rho _{0}) }{1+2\lambda _{e}k} dk     \,\,\, .
\label{bubble-annular}
\end{equation}
As and example, Fig. \ref{f4} displays the force when $M_{1}=|M_{2}|$,
$W_{1}=0.5\lambda_{e} $, $W_{3}=2W_{1}$, and $W_{2}=1.5W_{1}$. The solid line
shows the case of a positive $M_{2}$, whereas the dashed line corresponds to
negative $M_{2}$.  A slowly moving vortex
approaching the annular structure from the outside will experience attraction if
$M_{2}>0$, and its equilibrium position is directly under the annulus. If $M_{2}$
suddenly switches sign, then the vortex is attracted to the center of the
annular structure (i.e. directly under the dot) and kept there. Thus, this 
annular structure could be used to catch and trap vortices by switching the
sign of the outer annulus. This could in principle be done by using a soft
magnetic material with at the outer annulus, whereas the
inner dot consists of a hard magnetic material. Naturally, the magnetic forces
must overcome the Lorentz forces generated by the external magnetic field.
Moreover, it is probably not so easy to fabricate such a structure.
Therefore, it is doubtful whether this circuit could find any practical
realization, although it is a nice example of a vortex manipulation structure.  

\section{Interaction with a 1D magnetization}
Most domains and walls are not cylindrical symmetric. In fact, they are more
often one dimensional (1D), and therefore the corresponding London equation
should be solved for this symmetry. Consider a 1D magnetization distribution 
with the vector pointing in the z direction. Then we may apply the methods 
presented in this article (using the 1D Fourier transform), and find the 
following expression for the scalar potential:
\begin{equation}
\phi _{m}(x,z) =-\frac{\lambda_{e}}{\pi }\int_{0}^{\infty} 
\frac{k_{x}M(k_{x})cos(k_{x}x)}{1+2\lambda _{e}k_{x}} \exp(-k_{x} |z|)dk_{x}
\,\,\, ,
\label{mag2}
\end{equation}
where we have assumed that the Fourier transform of the magnetization is real.
The magnetic field components are given by
\begin{equation}
H_{mz}(x,z) =\frac{\lambda _{e}}{\pi }\int_{0}^{\infty}k_{x}^{2} \frac{M(k_{x}) cos(k_{x}x)}{1+2\lambda _{e}k_{x}}
\exp(-k_{x}|z|)dk_{x} \,\,\, ,
\end{equation}
\begin{equation}
H_{mx}(x,z) =\frac{\lambda _{e}}{\pi }\int_{0}^{\infty}k_{x}^{2} \frac{M(k_{x}) sin(k_{x}x)}{1+2\lambda _{e}k_{x}}
\exp(-k_{x}|z|)dk_{x}
\,\,\, .
\end{equation}
The energy of the system is found to be
\begin{equation}
E = \frac{\Phi _{0} ^{2}}{4\pi \mu _{0} \lambda _{e}} ln\left( \frac{4\pi
\lambda _{e}}{\xi} \right) - \frac{\Phi_{0}}{\pi} \int_{0}^{\infty} \frac{M(k_{x}) 
cos(k_{x}x_{0})}{1+2\lambda _{e}k_{x}} dk_{x}  - \frac{\lambda _{e}\mu _{0}}{2\pi} \int_{0}^{\infty} 
k_{x}^{2} \frac{M^{2}(k_{x})}{1+2\lambda _{e}k_{x}} dk_{x} \,\,\, .
\end{equation}
To find the interaction force between the magnetization distribution and the 
vortex, we simply take the negative derivative of the interaction energy, 
and obtain
\begin{equation}
F (x_{0}) =  -\frac{\Phi _{0}}{\pi} \int_{0}^{\infty} k_{x}\frac{M(k_{x})
sin(k_{x}x_{0})}{1+2\lambda _{e}k_{x}} dk_{x}     \,\,\, .
\label{Bloch-vortex}
\end{equation}

As an example, Fig. \ref{f5} shows the z component of the magnetic field
generated by the following magnetization distribution (which is looks like a Bloch
wall separated by in-plane magnetized domains):
\begin{displaymath}
\mbox{\boldmath $M$} = \left\{ \begin{array}{ll}
M_{0}\delta(z)\mbox{\boldmath $\hat{e}$}_{z} & \textrm{if  $-W \leq x \leq
W$}\\
0 & \textrm{if $|x|  \geq W$}\\  
\end{array} \right. .
\end{displaymath}
Here we have chosen W=$\lambda_{e} /50$, z=0 and t=$\lambda_{e} /2000$. 
Note in particular the peaks near the edges of the Bloch
wall. In general, it is also seen that there is a small field directly above the
magnetization distribution which is not screened by the superconductor. It can
be shown that the corresponding force on the vortex has a shape similar to 
that of Fig. \ref{f2}, with its maximum near the edges of the magnetization
distribution.

\section{Conclusion}
The interplay between magnetic and superconductive films is studied. The
generalized London equation for this system is solved, and the magnetic
fields, the energy and the interaction forces are computed. In
particular, we focus on how to manipulate vortices using  
magnetic nanostructures. The formalism presented here could be generalized to
arbitrary magnetization vectors, and we therefore hope that it will be useful in
future studies.

\begin{figure}
\caption{The system under study. The thin oxide layer separating the the
superconductive and magnetic films is not shown. 
\label{f1}}
\end{figure}

\begin{figure}
\caption{The force between a vortex and a bubble when W=$\lambda_{e}$.  
\label{f2}}
\end{figure}

\begin{figure}
\caption{An annular structure for manipulating of vortices.  
\label{f3}}
\end{figure}

\begin{figure}
\caption{The force on a vortex near the annular structure of Fig. \ref{f3}.  
\label{f4}}
\end{figure}

\begin{figure}
\caption{The magnetic field generated by a Bloch wall with width
0.02$\lambda_{e}$. Here we have assumed that z=0.0005$\lambda_{e}$.  
\label{f5}}
\end{figure}

\newpage
\centerline{\includegraphics[width=12cm]{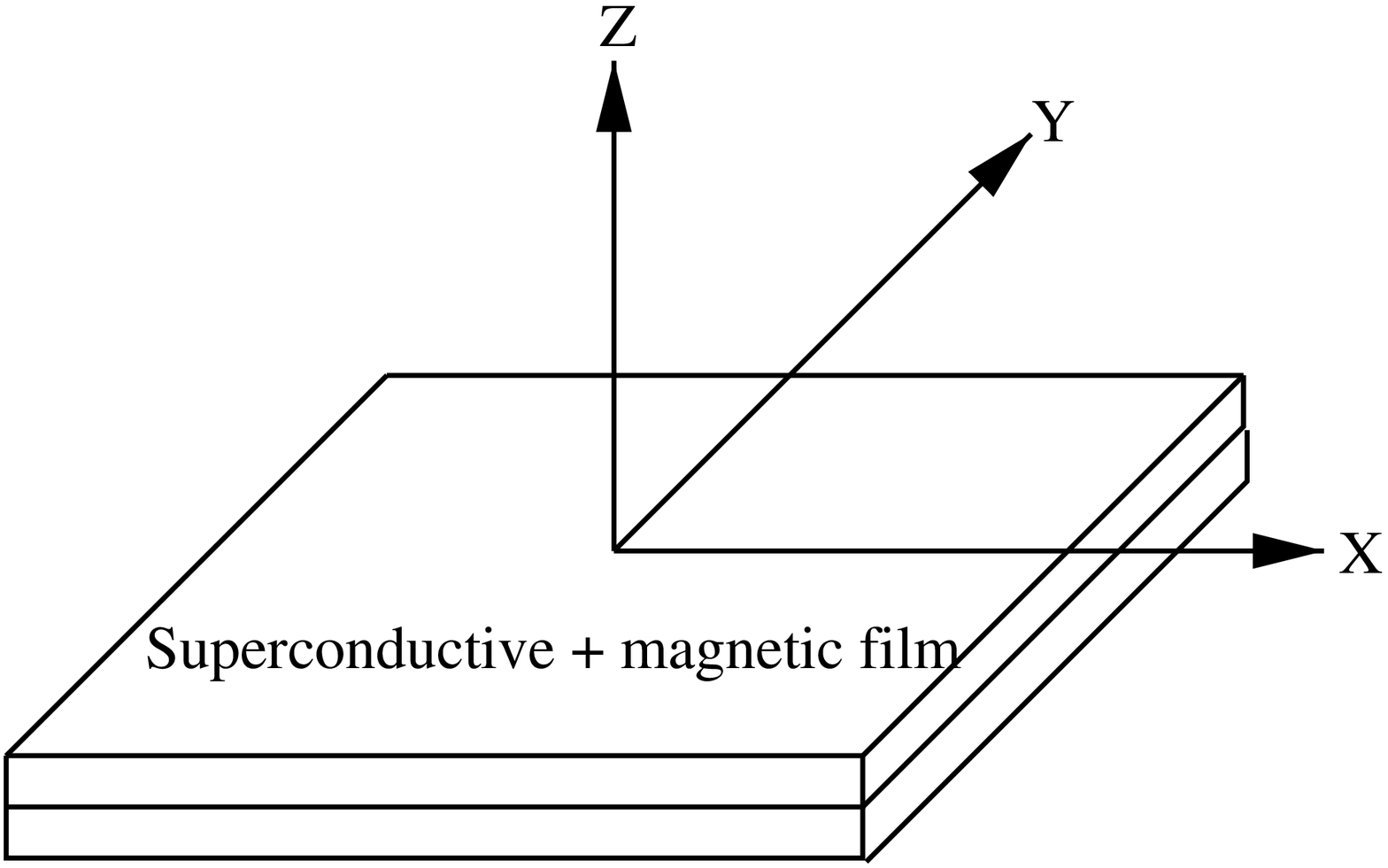}}
\vspace{2cm}
\centerline{Figure~\ref{f1}}

\newpage
\centerline{\includegraphics[width=12cm]{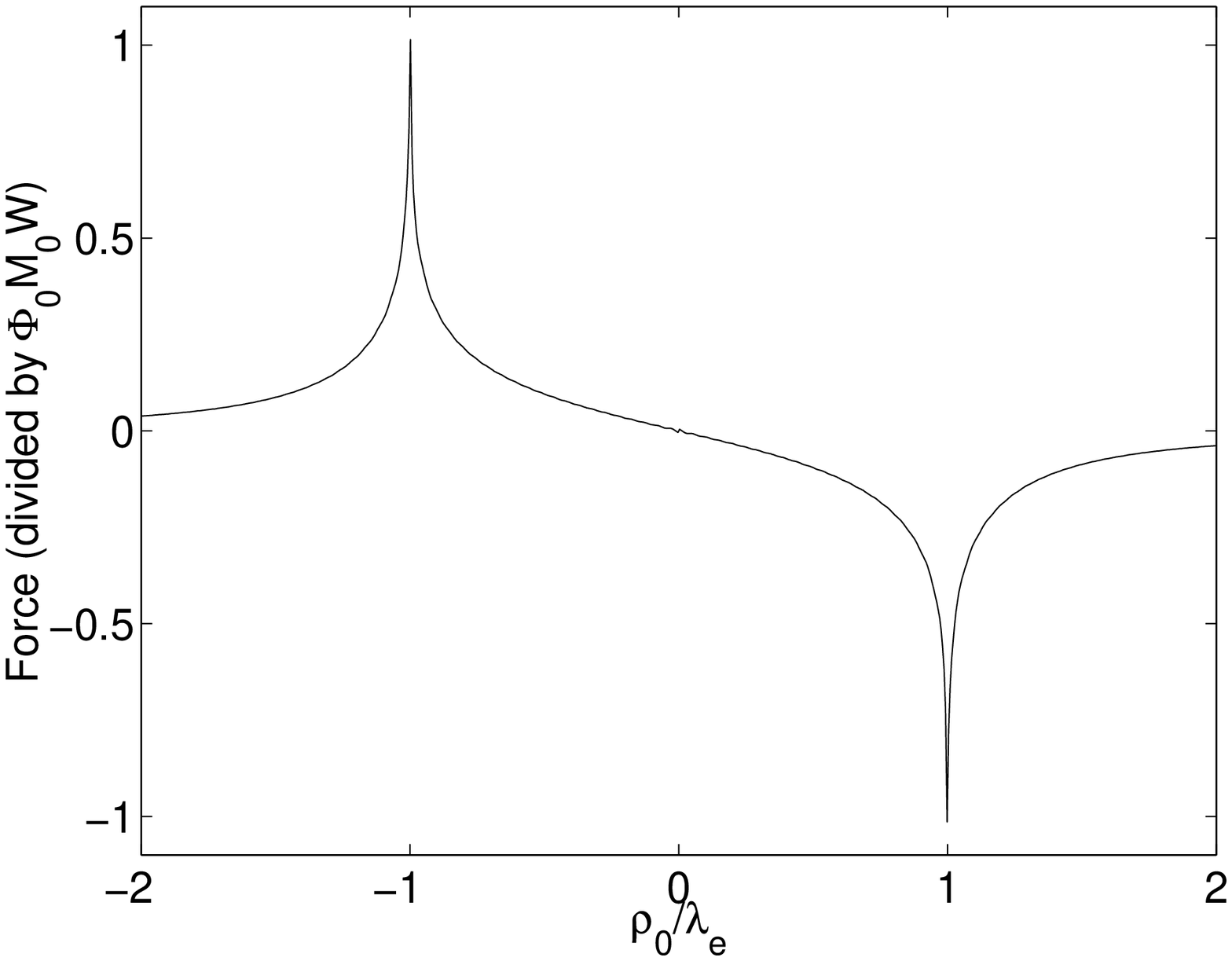}}
\vspace{2cm}
\centerline{Figure~\ref{f2}}

\newpage
\centerline{\includegraphics[width=12cm]{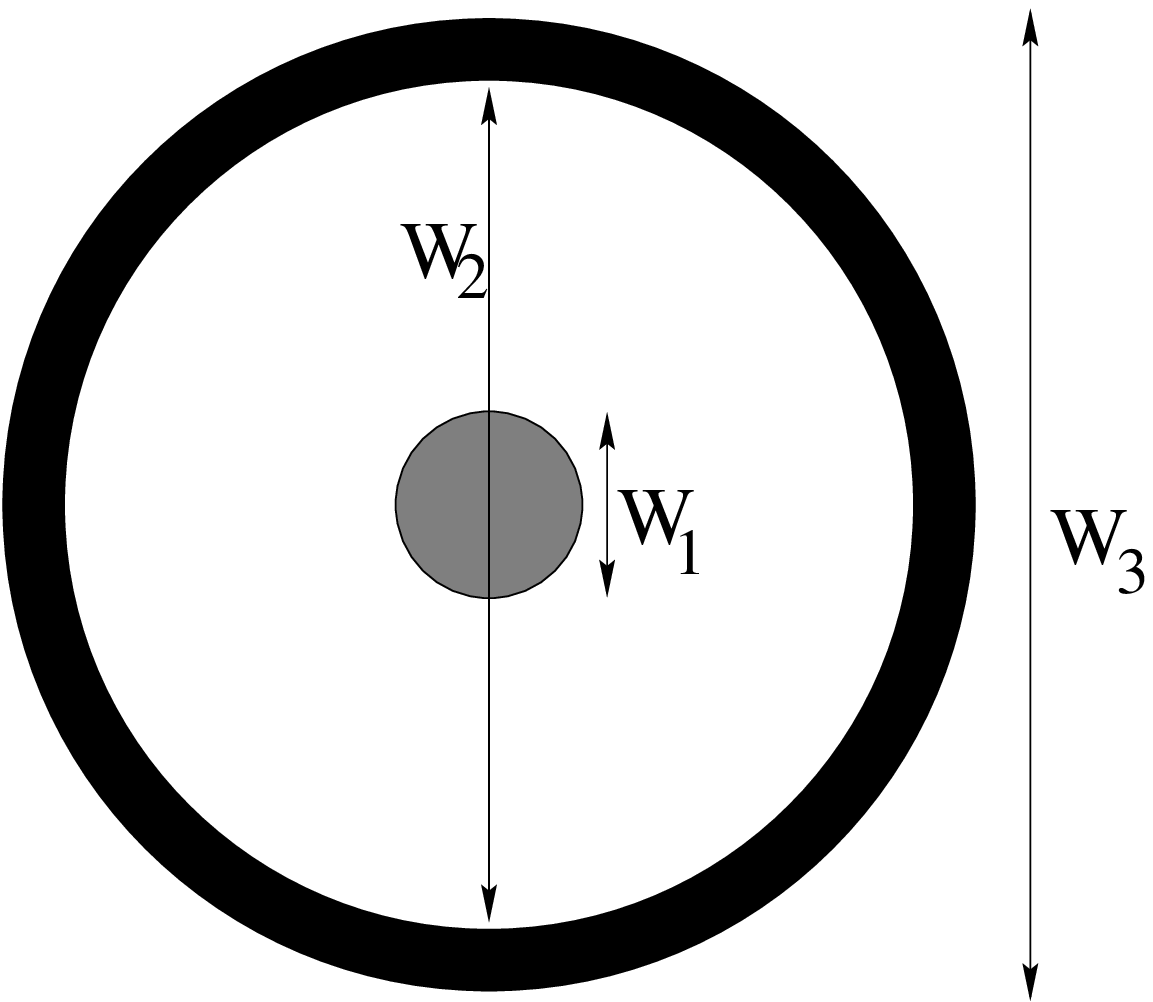}}
\vspace{2cm}
\centerline{Figure~\ref{f3}}

\newpage
\centerline{\includegraphics[width=12cm]{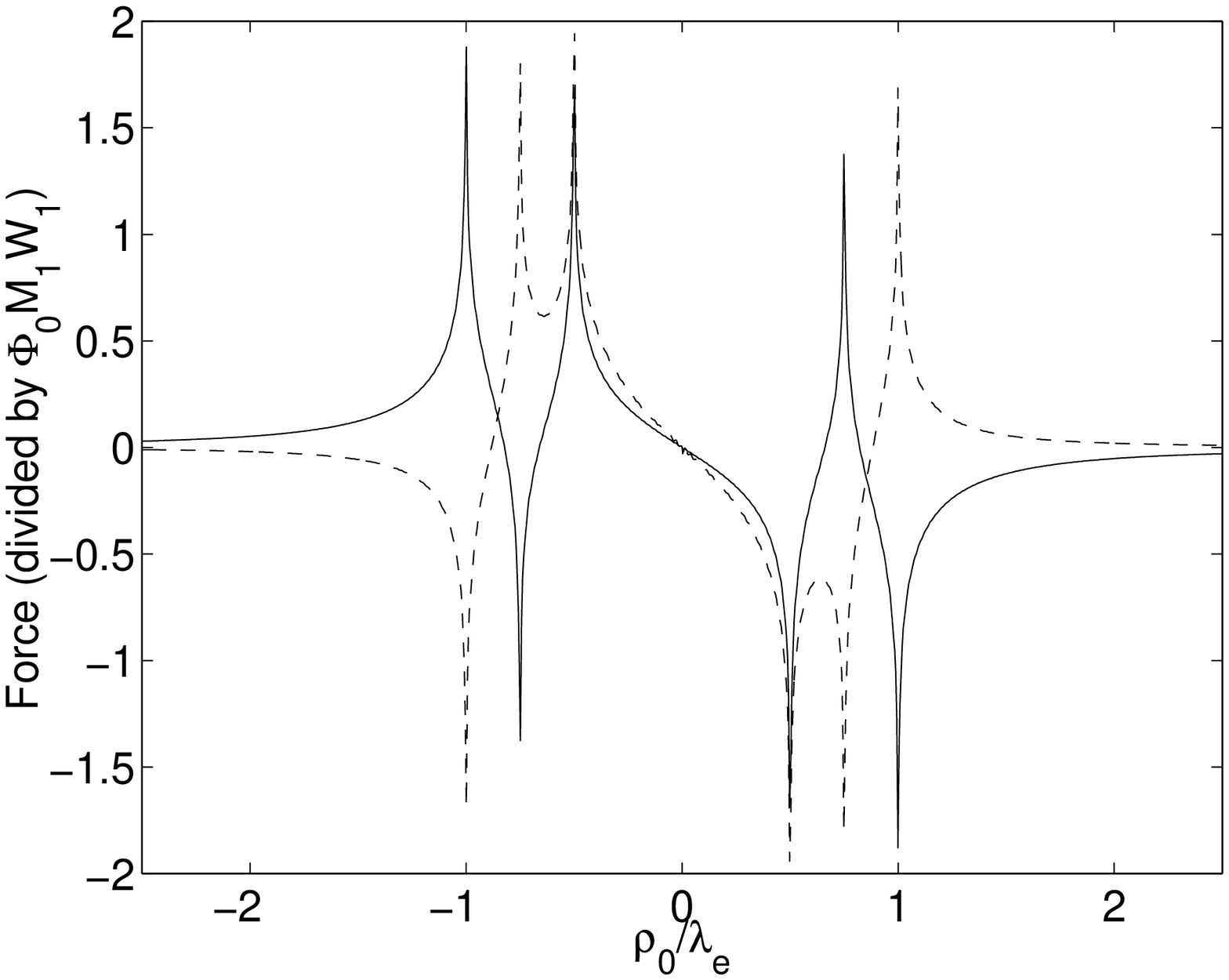}}
\vspace{2cm}
\centerline{Figure~\ref{f4}}

\newpage
\centerline{\includegraphics[width=12cm]{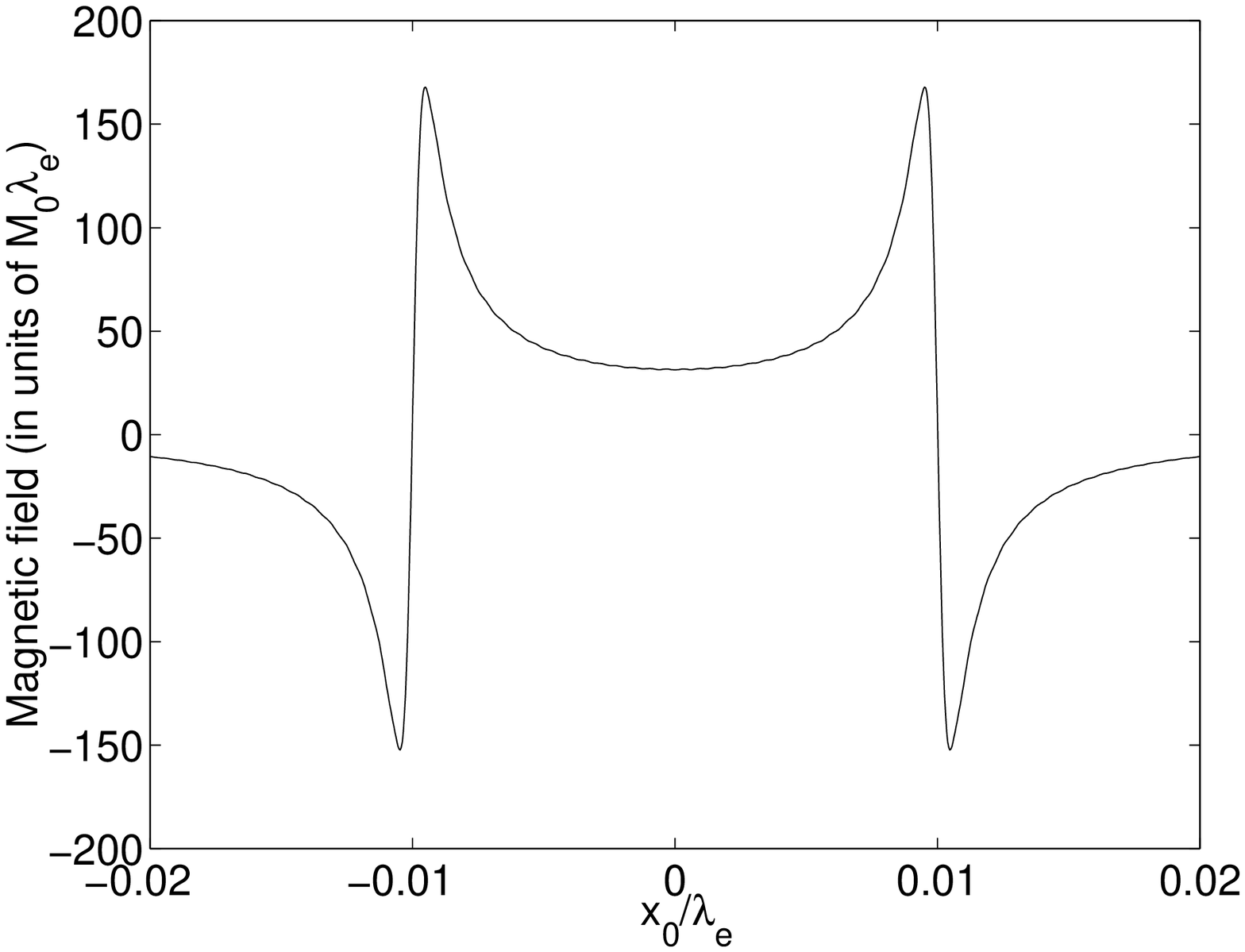}}
\vspace{2cm}
\centerline{Figure~\ref{f5}}

\end{document}